\documentclass[iop]{emulateapj}
\include{amssymb}

\begin{document}
Accepted to be published in ApJ

\title{THE ACS LCID PROJECT. IX. IMPRINTS OF THE EARLY UNIVERSE IN THE RADIAL VARIATION OF THE STAR FORMATION HISTORY OF DWARF GALAXIES.\altaffilmark{1}}

\author{Sebastian L. Hidalgo\altaffilmark{2,3},
Matteo Monelli\altaffilmark{2,3},
Antonio Aparicio\altaffilmark{2,3},
Carme Gallart\altaffilmark{2,3},
Evan D. Skillman\altaffilmark{4},
Santi Cassisi\altaffilmark{5},
Edouard J. Bernard\altaffilmark{6},
Lucio Mayer\altaffilmark{7},
Peter Stetson\altaffilmark{8},
Andrew Cole\altaffilmark{9}, and
Andrew Dolphin\altaffilmark{10}
}

\altaffiltext{1}{Based on observations made with the NASA/ESA Hubble Space Telescope,
obtained at the Space Telescope Science Institute, which is operated by the
Association of Universities for Research in Astronomy, Inc., under NASA contract NAS
5-26555. These observations are associated with program \#10505}
\altaffiltext{2}{Instituto de Astrof\'\i sica de Canarias. V\'\i a L\'actea s/n.
E38200 - La Laguna, Tenerife, Canary Islands, Spain;
shidalgo@iac.es, monelli@iac.es, aparicio@iac.es, carme@iac.es}
\altaffiltext{3}{Department of Astrophysics, University of La Laguna. V\'\i a L\'actea s/n.
E38200 - La Laguna, Tenerife, Canary Islands, Spain}
\altaffiltext{4}{Minnesota Institute for Astrophysics, University of Minnesota, Minneapolis, MN 55455, USA;
skillman@astro.umn.edu}
\altaffiltext{5}{INAF-Osservatorio Astronomico di Collurania,
Teramo, Italy; cassisi@oa-teramo.inaf.it}
\altaffiltext{6}{Institute for Astronomy, University of Edinburgh,
Royal Observatory, Blackford Hill, Edinburgh EH9 3HJ, UK; ejb@roe.ac.uk}
\altaffiltext{7}{Department of Physics, Institut f\"ur Astronomie,
ETH Z\"urich, Z\"urich, Switzerland; lucio@phys.ethz.ch}
\altaffiltext{8}{Dominion Astrophysical Observatory, Herzberg Institute of
Astrophysics, National Research Council, 5071 West Saanich Road, Victoria,
British Columbia V9E 2E7, Canada; peter.stetson@nrc-cnrc.gc.ca}
\altaffiltext{9}{School of Mathematics \& Physics, University of Tasmania,
Hobart, Tasmania, Australia; andrew.cole@utas.edu.au}
\altaffiltext{10}{Raytheon; 1151 E. Hermans Rd., Tucson, AZ 85706, USA;
adolphin@raytheon.com}

\begin{abstract}

Based on Hubble Space Telescope observations from the {\it Local Cosmology from Isolated Dwarfs} project, we present the star formation histories, as a function of galactocentric radius, of four isolated Local Group dwarf galaxies: two dSph galaxies, Cetus and Tucana, and two transition galaxies (dTrs), LGS-3 and Phoenix. The oldest stellar populations of the dSphs and dTrs are, within the uncertainties, coeval ($\sim 13$~Gyr) at all galactocentric radii. We find that there are no significative differences between the four galaxies in the fundamental properties (such as the normalized star formation rate or age-metallicity relation) of their outer regions (radii greater than four exponential scale lengths); at large radii, these galaxies consist exclusively of old ($\gtrsim 10.5$~Gyr) metal-poor stars. The duration of star formation in the inner regions vary from galaxy to galaxy, and the extended central star formation in the dTrs produces the dichotomy between dSph and dTr galaxy types. The dTr galaxies 
show prominent radial stellar population gradients: the centers of these galaxies host young ($\lesssim 1$~Gyr) populations while the age of the last formation event increases smoothly with increasing radius. This contrasts with the two dSph galaxies. Tucana shows a similar, but milder, gradient, but no gradient in age is detected Cetus. For the three galaxies with significant stellar population gradients, the exponential scale length decreases with time. These results are in agreement with outside-in scenarios of dwarf galaxy evolution, in which a quenching of the star formation toward the center occurs as the galaxy runs out of gas in the outskirts.

\end{abstract}


\section{INTRODUCTION}\label{secint}

Stellar population gradients in galaxies are important tracers of the physical processes shaping their structure and evolution. The nearest galaxies, for which quantitative star formation histories (SFHs) can be derived, are ideal candidates for a comparison with models studying the spatially resolved evolution of the stellar populations in galaxies. In this case, the analysis of deep color-magnitude diagrams (CMDs) (ideally reaching the oldest main sequence turnoffs, oMSTOs) and/or spectroscopy of individual stars, provides solid constraints on the structural evolution of the galaxy.

The study of our nearest neighbors, the Milky Way satellites, has revealed the presence of metallicity \citep{tolstoyetal2004, battagliaetal2006, carreraetal2008a, carreraetal2008b, carreraetal2011, battagliaetal2011, kirbyetal2011} and/or star formation history gradients \citep{carraroetal2002, gallartetal2008, leeetal2009, deboeretal2012} in most studied galaxies. Gradients are in the sense that stars are, on average, more metal poor and older outwards. These trends are qualitatively well reproduced by most models which study dwarf galaxy population gradients available in the literature (\citealp{stinsonetal2009, schronyenetal2011, revaz&jablonka2012}; \citealp[but see][]{schronyenetal2013}). The cause of the gradients build-up is attributed to a gradual concentration of star formation toward the central parts of the galaxy, where, therefore most of the metal enrichment also takes place. Stellar migration is found to be not playing a significant role neither in increasing or erasing the radial gradients 
originated at star formation \citep{stinsonetal2009, schronyenetal2013}.

Milky Way satellites, however, are expected to have suffered the effects of repeated strong interactions with their host. The morphological transformation of dSph from gas-rich, star forming systems has been linked to the effects of these interactions \citep[see][for a review]{mayer2010}, which are expected also to modify the distribution of their stellar populations \citep{lokasetal2012}. For this reason, it is important to characterize the stellar population gradients in isolated dwarf galaxies, in which their build-up has presumably been less complicated by external factors (in fact most models referenced above study isolated galaxies only).  

This paper is part of the Local Cosmology from Isolated Dwarfs (LCID) project series of papers. The aim of this project is to obtain the detailed SFHs of six isolated dwarf galaxies of the Local Group: Leo-A, Phoenix, Cetus, Tucana, LGS-3, and IC1613, in order to shed light on the effects of phenomena that may affect the early evolution of dwarf galaxies such as cosmic UV-background and internal feedback, in the absence of strong environmental effects. The global SFHs of these galaxies have been described in \citet{coleetal2007}, \citet{hidalgoetal2009}, \citet{monellietal2010a,monellietal2010b}, \citet{hidalgoetal2011}, and Skillman et al. (2013, in prep.), respectively. The results presented in these papers describe the full SFHs of the observed fields of the galaxies but, in all the cases but Phoenix, they lack a detailed  study of the SFHs as a function of the galactocentric radius. This paper will discuss the galactocentric variations of the SFH in the two dSph galaxies, Cetus and Tucana, and the two 
transition (dTr)\footnote{Note that transition galaxies, which are galaxies with relatively recent star formation but no prominent \ion{H}{2} regions, are often labeled as dIrr/dSph galaxies \citep[e.g.,][]{mateo1998}.  Here we will use dTr, which is simpler.}  galaxies, LGS-3 and Phoenix. A forthcoming paper will analyze the SFH gradients of the dIrr Leo A (Hidalgo et al. 2013, in prep).

In the context of the LCID project, there is an additional issue that we wish to investigate, namely, the possible effects of self-shielding in shaping up the galaxy's stellar populations. Self-shielding has been proposed as a mechanism to maintain the star formation in the inner regions of dwarf galaxies, protecting the gas in these denser regions from being heated by the UV-background \citep{susa&umemura2004,sawalaetal2010}. In this way, the effects of the UV-background could be best appreciated in the external parts of galaxies. A main result of the LCID project is that UV-background {\it alone} has not been instrumental in halting star formation in even the LCID dwarfs with less extended SFH \citep{hidalgo2011, monellietal2010a,monellietal2010b, hidalgoetal2011}, but a detailed radial analysis is required if we want to extend the former result to any galactocentric distance.

The organization of this paper is as follows. In \S\ref{secdata}, the data and spatial sampling is presented. The SFHs of the galaxies are compared as a function of galactocentric distance in \S\ref{secsfh}. The age-metallicity relation and SFHs as a function of galactocentric radius are presented in \S\ref{secamr} and \S\ref{secrad}, respectively. An analysis of the radial distribution of the stellar populations is performed in \S\ref{secgro}. The main conclusions of the work are summarized in \S\ref{secsum}.

\section{DATA AND SPATIAL SAMPLING}\label{secdata}
                                                                           
Detailed descriptions of the LCID observations of Phoenix, LGS-3, Cetus, and Tucana were presented in \citet{hidalgoetal2009,hidalgoetal2011} and \citet{monellietal2010a,monellietal2010b}, respectively. The reader is referred to these papers for a description of the procedures adopted to obtain the photometry and the false stars tests. Here we describe the procedure used to divide the observed fields in elliptical regions to obtain the SFHs as a function of galactocentric radius.

To sample the observed fields of LGS-3, Cetus, and Tucana, we have used, for each galaxy, an inner elliptical region and a set of partially overlapping elliptical annuli with increasing galactocentric radius. Both the inner elliptical region and the elliptical annuli are concentric and centered in the center of the galaxy and contain about 25\% of the total observed stars each. The reason for these criteria is twofold. On the one hand, by using overlapping regions the spatial sampling is increased and statistical fluctuations are smoothed out. On the other hand, we have checked  \citep{hidalgoetal2011} that 25\% is about the minimum number of stars to be used in order to obtain solutions for the SFH not significantly different from the original one, based on the total available stars. In total, 13 regions have been used for Cetus, 7 for Tucana and 10 for LGS-3. For Phoenix, we have selected five of the ellipses shown in Fig. 11 of \citet{hidalgoetal2009} which are based on \citet{martinezdelgado1999}.
  
To create the elliptical regions described above, the following procedure was used. For LGS-3 and Tucana, the stars included in the photometry lists were injected into an empty image using IRAF and the coordinates and magnitudes in the F814W filter. The resulting synthetic image was filtered with a two-dimensional rotationally symmetric Gaussian filter with a standard deviation of 100 pixels (4.9 arcseconds) and a kernel size of 500 pixels. This process suppresses noise and fluctuations on small scales in the image. Isophotes were determined in the filtered image using the {\it contour} function in $\rm Matlab^{\circledR}$. Ellipses were fitted to the contours given by the isophotes by using the Matlab function {\it ellipsefit} \citep{halivr&flusser1998}. The free geometric parameters of the ellipses used in the fit were the center, semi-major axis, ellipticity, and position angle. After the fit, the mean center of all the ellipses was calculated and all the ellipses were shifted to this common center 
position. In the case of Cetus, for which the center of the galaxy was not in our fields, the geometric parameters from \citet{mcconnachie&irwin2006} were used to set the center, ellipticity and position angle of the ellipses.

\section{THE CMD MORPHOLOGIES AND SFHs}\label{secsfh}

\begin{figure}
\centering
\includegraphics[width=9cm,angle=0]{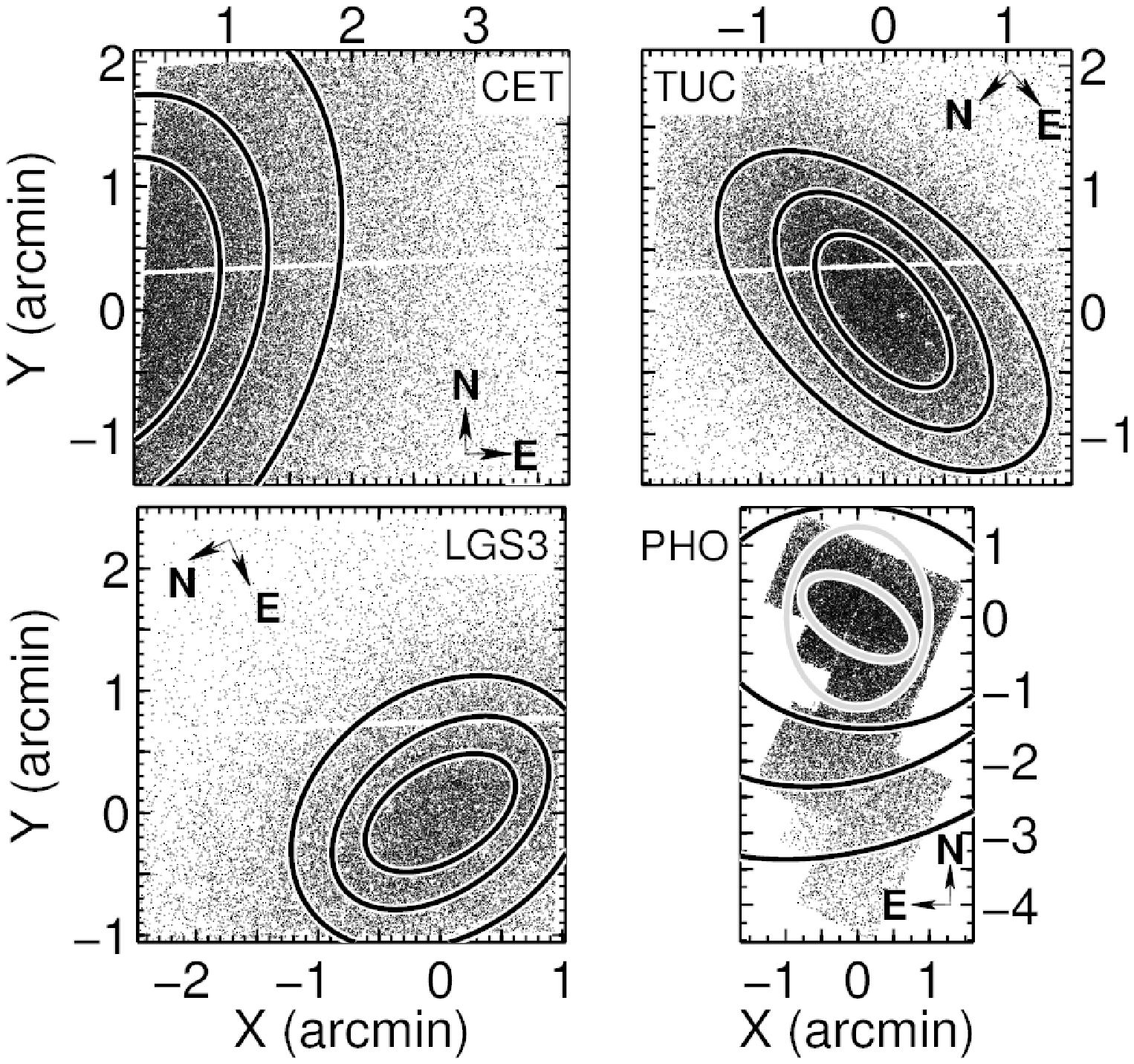}
\protect\caption[ ]{Regions $R_1$ to $R_4$ (form inner to outer) used to compare the CMDs and SFHs of the galaxies. Except for Phoenix, $R_1$ to $R_4$ contain approximately the same number of stars in each galaxy. For Phoenix two additional ellipses, in gray line, have been used to obtain the radial distribution of the stars. The origin of the axis corresponds to the center of the innermost ellipse. Field orientation is marked.}\label{f1}
\end{figure}

In this section, we will compare the CMDs and SFHs of the galaxies as a function of galactocentric radius. For simplicity, we will focus on four specific, non overlapping regions selected from the sample defined above. Except for Phoenix, the equivalent radii for the inner three regions approximately correspond to $\alpha_\psi$, $1.5\alpha_\psi$, and $2\alpha_\psi$ (where $\alpha_\psi$ is the exponential scale length of the galaxy's stellar mass distribution). The fourth region corresponds to the area outside of $\sim 2\alpha_\psi$. Hereafter, we will denote these regions, from inner to outer, as $R_1$, $R_2$, $R_3$, and $R_4$. For Phoenix, the three outermost ellipses have been re-scaled to sample similar radial distances to $R_1$, $R_2$, and $R_3$. We have not used $R_4$ in Phoenix due to the low number of observed stars (less than 3\% of the total). The limits of $R_1$ to $R_4$ are marked in Figure \ref{f1}. Table \ref{t1} gives the size in pc of each region (equivalent radius) and the number of stars 
within each region.

\begin{figure}[h]
\centering
\includegraphics[width=9cm,angle=0]{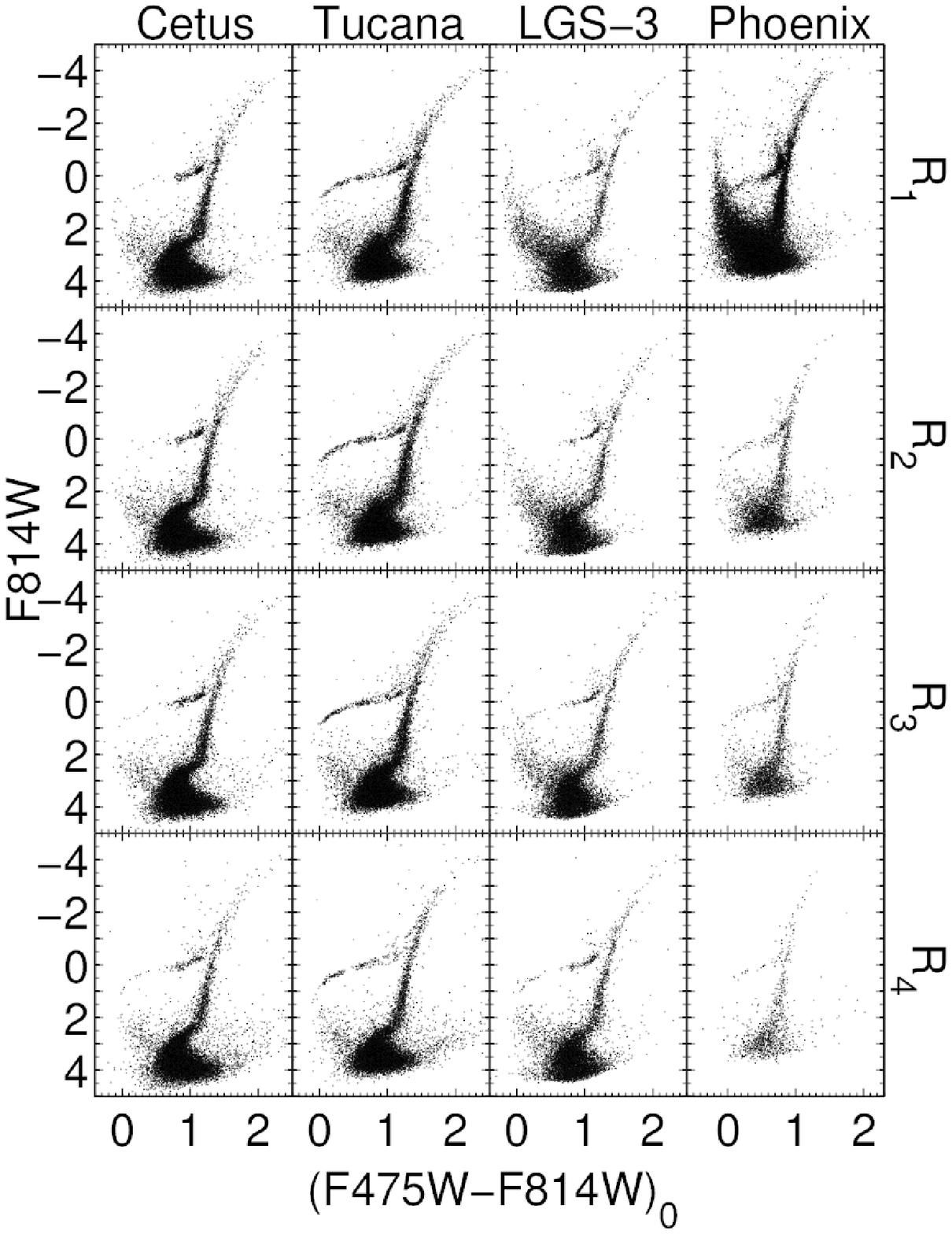}
\protect\caption[ ]{Color-magnitude diagrams for Cetus, Tucana, LGS-3 and Phoenix for the stars located
in the regions marked on the right axis. For the case of Phoenix, the color and magnitude axes are actually V-I and I, respectively. The location of $R_1$ to $R_4$ is shown in Figure \ref{f1}. The distance to the galaxy center increases from top to bottom.}\label{f2}
\end{figure}

\begin{deluxetable*}{lcccc}[t]
\tabletypesize{\scriptsize}
\tablecaption{Limits for elliptical regions and number of stars within.\label{t1}}
\tablewidth{0pt}
\tablehead{
\colhead{Galaxy}       &\colhead{$R_1$}                     &\colhead{$R_2$}  &\colhead{$R_3$} &\colhead{$R_4$}\\
\colhead{}  &\colhead{(pc , (\#stars))}     &\colhead{(pc , (\#stars))}              &\colhead{(pc , (\#stars))}      &\colhead{(pc , (\#stars))}} 
\startdata
CETUS\tablenotemark{a}      &0\,--\,242.9 (12948)           &242.9\,--\,342.0 (13053)                &342.0\,--\,494.2 (13722)        &494.2\,--\,1085.6 (12263)\\
TUCANA                      &0\,--\,137.6   (13323)         &137.6\,--\,218.4 (13970)                &218.4\,--\,321.1 (14414)        &321.1\,--\,\,\,\,817.7  (11939)\\
LGS-3                       &0\,--\,\,\,\,98.1\,\,\, (7461) &\,\,\,98.1\,--\,152.5\,\,\, (7419)      &152.5\,--\,219.2\,\,\, (7471)   &219.2\,--\,\,\,\,678.3\,\,\,  (7473)\\
PHOENIX                     &0\,--\,220.9 (26388)           &220.9\,--\,314.2\,\,\, (3289)           &314.2\,--\,448.9\,\,\, (2525)   &448.9\,--\,\,\,\,628.5\,\,\,  (1042)\\
\enddata
\tablenotetext{a}{We have assumed that the SFH of the inner ellipse obtained for Cetus is representative of the SFH in the center of the galaxy which is outside the observed field.}
\end{deluxetable*}

Figure \ref{f2} shows the CMDs of Cetus, Tucana, LGS-3, and Phoenix for each region. The CMDs of the dTrs, LGS-3 and Phoenix, show a gradual decrease in the number of stars located in the blue-plume with increasing distance from the center. The CMD morphologies of the blue-plumes of the dSphs, Cetus and Tucana, show no significant change with radius. \citet{monellietal2012b} have identified these stars as a population of blue-stragglers (BSSs) with a flat radial distribution. On another hand, \citet{bernardetal2009} have pointed to a change in the morphology of the horizontal-branch (HB) and red giant branch (RGB) with radius in these two galaxies, which can also be observed in the Fig. \ref{f2} (particularly that of the HB). It is interesting to note that the CMD of LGS-3 at $R_4$ is remarkably similar to those of Cetus and Tucana at all radii. These changes in the morphologies of the CMDs indicate a radial 
gradient of the stellar populations, more evident in LGS-3 and Phoenix.

Using the CMDs shown in Fig. \ref{f2} we have obtained the SFHs of Cetus, Tucana, LGS-3, and Phoenix as a function of radius. We have used IAC-star/IAC-pop/MinnIAC to obtain the SFHs as described in \citet{aparicio&hidalgo2009} and \citet{hidalgoetal2011}.  In short, IAC-star \citep{aparicio&gallart2004} is used to create a synthetic CMD, whose distribution of stars is compared with that in the observed CMD, after suitable simulation of observational effects, which include the dependency with the position of the star in the field. This comparison is carried out using the IAC-pop/MinnIAC algorithms which produce the SFH: mass of gas converted into stars in the galaxy as a function of age and metallicity. This technique does not assume any age-metallicity relation (AMR) for the stars and explores possible variations in the values of the photometric zero points, distance modulus, and reddening, thus minimizing the impact of the uncertainties in these parameters on the final SFH. In general, we denote the 
SFH defined above as $\psi(t,Z)$, while we define the star formation rate (SFR) as the SFH as a function of time only, i.e., $\psi(t)$. In \S\ref{secrad} we introduce also the dependency of the SFR with radius, $\psi(t,r)$.

To obtain $\psi(t,Z)$ for the regions described above, we have used the functions and parameters adopted in \citet{hidalgoetal2009}, \citet{hidalgoetal2011}, \citet{monellietal2010a}, and \citet{monellietal2010b}. In short, the synthetic CMD contains $8\times 10^6$ stars generated assuming a constant SFR in the range of age 0\,--13.5 Gyr and a metallicity range Z = 0.0001--0.005. The CMDs features sampled in both the synthetic and the observed CMDs are the main-sequence and the subgiant stars. Neither the RGB nor the HB stars were used to obtain the solution. The reader is referred to the former references for a full description of the method and the constrains adopted.

\begin{figure}[h]
\centering
\includegraphics[width=9cm,angle=0]{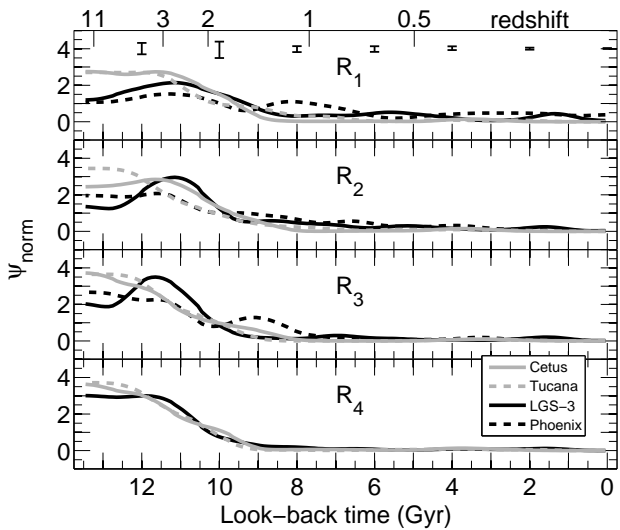}
\protect\caption[ ]{Normalized SFRs obtained using the CMDs shown in Figure \ref{f2} and the software suite IAC-star/IAC-pop/MinnIAC. The SFR of Phoenix has not been obtained for $R_4$ due to poor statistics. The SFRs have been normalized to their time integral for each region and galaxy. Error bars shown in the top panel are typical $1\sigma$ confidence intervals of the solutions at any radii. The redshift scale on the top has been obtained using a flat Einstein-de Sitter universe and the 5-year WMAP data \citep{komatsuetal2009}.}\label{f3}
\end{figure}

Figure \ref{f3} shows the $\psi(t)$ for Cetus, Tucana, LGS-3, and Phoenix obtained for the four regions defined above. In the case of Phoenix, the $\psi(t)$ has been obtained only for the three innermost regions. The $\psi(t)$ has been normalized to its time integral for each region. Regardless of the distance to the center, there is star formation activity in all galaxies for ages $\gtrsim 10$~Gyr. In fact, in all the galaxies, the bulk of the star formation occurs before 10 Gyr ago at all radii. However, $\psi(t)$  gradually decreases outwards for ages $\lesssim 9$~Gyr. Interestingly, in the outermost area ($R_4$) the three normalized $\psi(t)$ are almost indistinguishable within error bars. The stellar populations observed at $\sim 3.5$~Gyr in Cetus and Tucana have been identified by \citet{monellietal2012b} as BSSs. We adopt this result here and therefore the conclusion that there is no star formation activity in these two galaxies in the last $6\sim 8$~Gyr. In the case of LGS-3, the expected BSSs may be 
mixed with some young and intermediate age stellar population, which may produce the bump at 1.5 Gyr in $\psi(t)$. In summary, the overall trend is that regardless the morphology of the galaxies, the main difference in their SFHs occurs in the inner regions, tending to disappear at larger distances from the center.

\section{THE AGE-METALLICITY RELATIONS}\label{secamr}

\begin{figure}[h]
\centering
\includegraphics[width=9cm,angle=0]{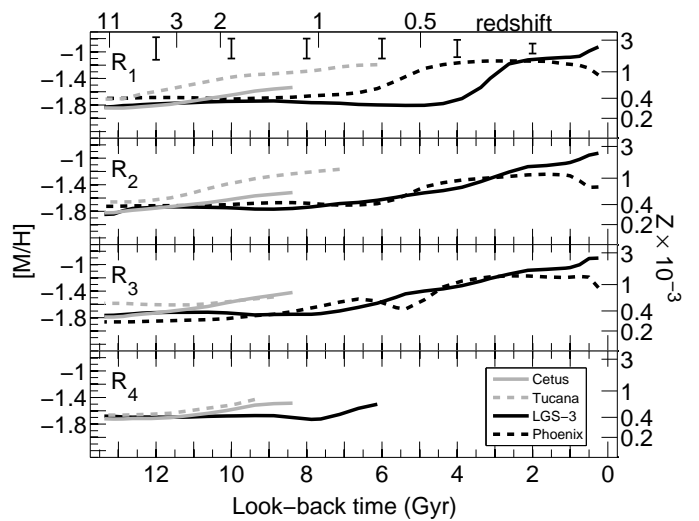}
\protect\caption[ ]{Age-metallicity relations for the galaxies of the sample. Right axis shows the metallicity $Z$. Left axis shows $\rm [M/H]$ (see $\S \ref{secamr}$). Error bars shown in the top panel are typical $1\sigma$ confidence intervals of the solutions at any radii.}\label{f4}
\end{figure}

Figure \ref{f4} shows the AMR for $R_1$ to $R_4$ in all the galaxies ($R_1$ to $R_3$ in the case of Phoenix). The overall trend is that metallicity increases with time. However, in the case of LGS-3 and Phoenix, the metallicity remains almost unchanged for most (LGS3) or half (Phoenix) of the lifetime of the galaxy in the central region. For these galaxies, the AMR relation can be separated into two distinct time periods: a first period involving the older stars in which the metallicity does not change significantly with time, and a final period where the metallicity increases more steeply. The duration of the first period appears to be a function of radius, decreasing with increasing radius. In contrast, the metallicities of Cetus and Tucana increase steadily with time, showing no trend with radius.

\begin{figure}[h]
\centering
\includegraphics[width=9cm,angle=0]{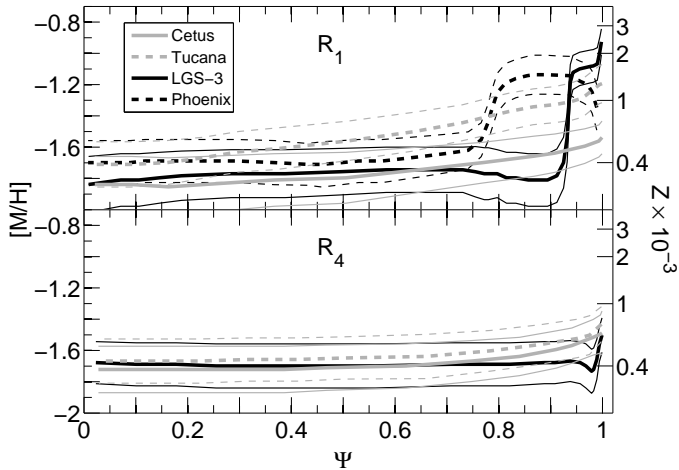}
\protect\caption[ ]{Metallicity as a function of the normalized cumulative star formation, $\Psi$, for all four galaxies (see text for the definition of $\Psi$). Right axis shows the metallicity $Z$. Left axis shows $\rm [M/H]$  (see $\S \ref{secamr}$).
Thin lines represent $1\sigma$ confidence intervals of each solution. Top and bottom panel show the innermost and outermost regions respectively.}\label{f5}
\end{figure}

To elucidate the nature of the two distinct time periods for metallicity enrichment in LGS-3 and Phoenix, we have plotted the metallicity of the stars formed at time $t$, $Z(t)$, as a function of the normalized total stellar mass formed between $t^\prime=0$ and $t^\prime=t$. That is, we use the bold-faced $\Psi(t) =\int_{0}^{t}\psi(t^\prime) dt^\prime / \int_{0}^{T}\psi(t^\prime) dt^\prime$ (where $T$ is the age of the Universe) as the normalized integrated SFR (as distinguished from the instantaneous SFR, $\psi(t)$). Figure \ref{f5} shows the metallicity as a function of $\Psi$ for the innermost ($R_1$) and outermost ($R_4$) regions for all galaxies of the sample. In the inner regions, LGS-3 and Phoenix show a sharp increase in the metallicity for $\Psi\gtrsim 0.90$ and $\Psi\gtrsim 0.75$, respectively. For smaller values of $\Psi$ there is almost no metallicity enrichment in LGS-3 or Phoenix. This effect is not observed for Cetus and Tucana, which show a steady increase of the metallicity with $\Psi$ in 
the innermost region. At $R_4$ a similar metallicity is inferred for all three galaxies at any $\Psi$. The outer region of LGS-3 does not show the steep increase of metallicity observed in its center.

\begin{figure}
\centering
\includegraphics[width=9cm,angle=0]{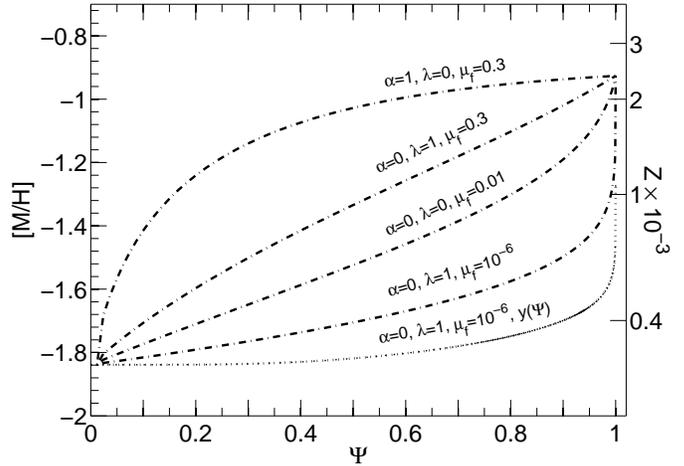}
\protect\caption[ ]{Metallicity enrichment laws computed using the SFR of the region $R_1$ of LGS-3. Closed-box ($\alpha=0, \lambda=0$),  infall ($\alpha=1, \lambda=0$) and outflow ($\alpha=0, \lambda=1$)
are shown for several final gas fractions $\mu_f$. An ad hoc, outflow model with a mass-dependent effective yield $y(\Psi)$ has been also computed.}\label{f6}
\end{figure}

To better understand the physical conditions implicit in the results shown in Fig. \ref{f5}, we have computed some simple chemical models. Figure \ref{f6} shows [M/H]\footnote{where we use the normal formalism of $\rm [M/H]=log(Z/Z_\sun)$} as a function of $\Psi$ calculated using the $\Psi$ of region $R_1$ of LGS-3 and different assumptions for the involved parameters and the chemical enrichment scenarios: closed-box, infall, and outflow. Here we have used the following notation: $\mu_f$ is the final gas fraction; $\alpha$ is the infall parameter, defined such that $f_I=\alpha(1-R)\psi(t)$ is the infall rate, where $R$ is the returned fraction; $\lambda$ is the outflow parameter, defined such that $f_O=\lambda(1-R)\psi(t)$ is the outflow rate \citep*[see][for a detailed description of these parameters]{peimbertetal1994}. The range of values tested for $\mu_f$ has been chosen to cover the range of current gas fractions of our galaxies, from no or very little gas in Cetus and Tucana to $\sim 0.15$ and $\sim 0.38$ in LGS-3 and Phoenix, respectively \citep{mcconnachie2012}. For simplicity, only a few of all the computed models are shown in Figure \ref{f6}. As indicated in the labels, one of the outflow models has been computed using a mass-dependent effective yield $y(\Psi)$. This ad hoc effective yield has been calculated such that the resulting AMR provides a closer match to the data for the inner parts of the dTrs. This condition is fulfilled with an effective yield in the form $y(\Psi)= Ae^{-1/\Psi}$, where $A$ is a constant. Given the characteristics of the SFHs of LGS-3 and Phoenix, which formed high fractions of their mass at the beginning of their evolution, this implies a fast chemical enrichment in the galaxy only when most of the stellar mass has already been formed.

\begin{figure}
\centering
\includegraphics[width=9cm,angle=0]{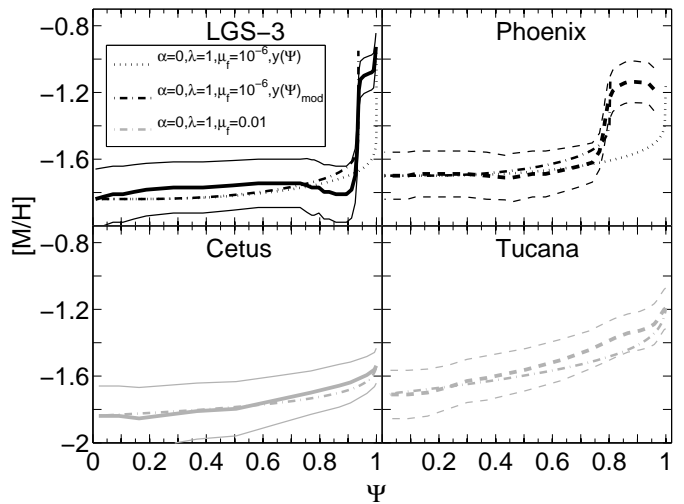}
\protect\caption[ ]{Metallicity as a function of $\Psi$ of all the galaxies (thick lines) compared with the optimal maching chemical models. Thin lines represent $1\sigma$ confidence intervals of each solution. Only the $R_1$ region is shown. The models which best match the observations of the dTrs are the ad hoc, outflow models with mass-dependent effective yield 
$y(\Psi)$ but for which most of the pristine gas is removed by $\Psi=0.90$ (LGS-3) and $\Psi=0.75$ (Phoenix), labeled $y(\Psi)_{mod}$. For the dSphs, simple outflow models provide adequate fits to the observations.}\label{f7}
\end{figure}

For the chemical models described above, we see that infall models fail for both dTrs and dSphs due to their characteristic fast enrichment at early times. Closed-box models do not allow significant star formation without concomitant metal enrichment, so only closed-box models with unrealistically low effective yield would mimic the observations. Outflow models allow for slow chemical enrichment and the possibility for more rapid enrichment at later times. Figure \ref{f7} shows optimal matching models for the $R_1$ region in each galaxy. In the cases of Cetus and Tucana, the outflow models ($\alpha=0, \lambda=1, \mu_f=0.01$) are in good agreement with the morphology of the $\Psi$-metallicity relation and also, qualitatively, with the conclusion by \citet{monellietal2010b} that gas loss connected to SNe events must have been very important in Tucana's early evolution\footnote{Note that for a fixed final gas fraction $\mu_f$, closed-box with effective yield and outflow models give similar AMRs \citep[see 
Fig. 10 in][]{hidalgoetal2009}. However, as discussed in \citet{monellietal2010b} the second is preferred.}  In contrast, for LGS-3 and Phoenix the outflow models with mass-dependent yield ($\alpha=0, \lambda=1, \mu_f=10^{-6}, y(\Psi)$) show good agreement for $\Psi\lesssim 0.80$ but not for larger values. In order to reproduce the results of LGS-3 and Phoenix for the full range of $\Psi$, a slightly modification of the mass-dependent yield model needs to be introduced: we impose that the remaining pristine gas is removed with the rapid increase of metallicity (i.e. for $\Psi=0.90$ and $0.75$ respectively).

\begin{deluxetable}{cccc}[h]
\tabletypesize{\scriptsize}
\tablecaption{Structural properties obtained from $\psi(r)$.\label{t2}}
\tablewidth{0pt}
\tablehead{
\colhead{Galaxy} &\colhead{$\alpha_\psi$}   &\colhead{$CR_\psi$}  &\colhead{$M_\star$}\\
\colhead{}       &\colhead{(pc)}            &\colhead{(pc)}       &\colhead{($10^6 M_\odot$)}}
\startdata
CETUS            &$212\pm 2 $               &$428\pm 2 $ &$7.0\pm 0.1$\\
TUCANA           &$121\pm 2 $               &$240\pm 2 $ &$3.2\pm 0.1$\\
LGS-3             &$111\pm 2 $               &$224\pm 3 $ &$1.9\pm 0.1$\\
PHOENIX          &$227\pm 10$               &$325\pm 15$ &$3.2\pm 0.3$\\   
\enddata
\end{deluxetable}

These simple chemical models provide some clues to the processes that may be responsible for the shapes of the $\Psi$-metallicity relations in the dTrs and their differences from the dSphs. The first phase of very little or no metal enrichment in LGS-3 and Phoenix accounts for 90\% and 75\% (respectively) of their total cumulative stellar mass formed and coincides with the old, main event of star formation in these galaxies, lasting 3 -- 4 Gyr. Most of the metal enrichment occurs in the second phase, that corresponds to 10\% (LGS-3) and 25\% (Phoenix) of the total stellar mass formed in these galaxies and in which $\psi(t)$ is much lower than during the first phase. The models are able to reproduce the sudden increase in the metallicity if an ad hoc effective yield $y(\Psi)$ is assumed and if most of the pristine gas has been lost from the galaxy by the end of the first phase (see $y(\Psi)_{mod}$ in Figure \ref{f7}).

All of this points to a scenario in which, in the first phase, most new formed metals are ejected from the galaxy without mixing with the interstellar medium. The early period characterized by a high $\psi(t)$ would sweep out most of the pristine gas and most of the metals produced in the galaxy. In the second phase, $\psi(t)$ would be so low that the number of stars able to inject the mechanical energy to produce a net outflow is very low, allowing all or a significant fraction of the enriched gas to remain in the galaxy. The stars born in this second phase would form from this enriched gas resulting in a rapid increase of the stellar metallicity. Only the central part of the galaxy had enough gas to carry on this process.

Interestingly, the effect of the negligible metallicity enrichment in LGS3 and Phoenix for most of their lifetimes can be observed in the width of the RGB. 
It is known that, for a fixed age range, the width of the RGB is indicative of the metallicity spread \citep*{gallartetal2005}. The main, initial episode of star formation in all galaxies of the sample accounts for more than 75\% of the total stellar mass and contains stars older than 8 Gyr. For this reason, a difference in the width of the RGB should be mainly related to differences in the metallicity distributions of the stars. As shown in Fig. \ref{f2}, the widths of the RGBs in the dTrs (measured as width at half maximum of the color distribution of the stars located between 1 and 1.5 mag below the red-clump) is narrower ($\sim 0.035$~mag) than in the dSphs ($\sim 0.050$~mag), as expected if most of the mass formed in the dTrs has been formed in the first phase with no metal enrichment.

\section{SFH AS A FUNCTION OF GALACTOCENTRIC RADIUS}\label{secrad}

\begin{figure}
\centering
\includegraphics[width=8cm,angle=0]{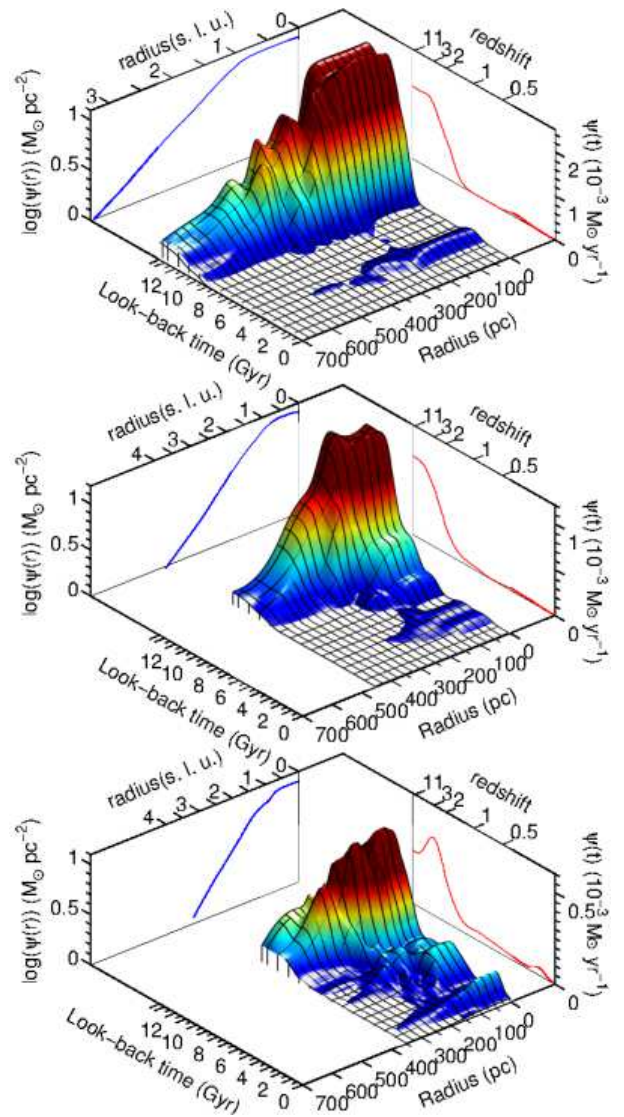}
\protect\caption[ ]{SFH of Cetus (top), Tucana (middle), and LGS-3 (bottom) as a function of time and galactocentric radius, $\psi(t,r)$. The normalized surface mass density of the total stars ever formed is shown as a function of time and radius. The SFR as a function of time, $\psi(t)$, is shown on the right, vertical panel. The density of the stellar mass, $\psi(r)$, is shown on the left vertical-plane as a function of radius.}\label{f8}
\end{figure}

Using the elliptical regions described in \S \ref{secsfh} (see Fig.~\ref{f1}), we have obtained the SFR as a function of age and galactocentric radius, $\psi(t,r)$, for the four galaxies. In the cases of Cetus, Tucana, and LGS-3, we have used a cubic spline data interpolation to obtain a continuous $\psi(t,r)$, which is shown over the age-radius axes in Figure \ref{f8}. The SFR as a function of time only, $\psi(t)= \int_{r=0}^{\infty}\psi(t,r) dr$, is shown on the right-hand, vertical axis. In all the cases, $\psi(t)$ is in full agreement with the global SFHs obtained for the individual galaxies shown in \citet{monellietal2010a,monellietal2010b}, and \citet{hidalgoetal2011}. On the left-hand vertical axis, the normalized surface mass density of the total stars ever formed, $\psi(r)=\int_{t=0}^T\psi(t,r) dt$, is shown as a function of galactocentric radius for each galaxy. In the case of Phoenix, $\psi(t,r)$ has been calculated for the five inner regions shown in Fig.~\ref{f1} with no interpolation.

The more detailed results of Cetus, Tucana, and LGS-3 allow to analyze their SFHs as a function of galactocentric distance, providing a deeper insight into their morphological evolution. The function $\psi(r)$ shows a two-component profile with a break radius $r_b$ located between 0.5 -- 1 scale lengths. In the three cases, the inner profile is almost flat within $r_b$. Most surface brightness profiles of dwarf disk galaxies show a two-component profile with the inner component flatter than the outer one \citep{hunter&elmegreen2006,zhangetal2012,herrmannetal2011}. Note that $\psi(r)$ is not necessarily the same as the surface brightness for two reasons: (i) because $\psi(r)$ refers to the total stars formed while only currently present stars contribute to the surface brightness and (ii) because the luminosity of young, bright stars can be an important contribution to the surface brightness profile and their luminosity changes rapidly with time. However, in this case we can assume that both are similar 
because the bulk of the star formation is located at intermediate and old ages and the surface brightness profile is not dominated by young stars.

Under some assumptions $\psi(r)$ can be used to obtain the total mass of stars ever formed in a galaxy even beyond the observed field. These assumptions are that (i) the observed field is a good sample of the rest of the galaxy and (ii) the radial profile of the stellar mass density follows the same $\psi(r)$ beyond the observed radius. For an exponential profile it can be represented as $\psi(r) = M_0 e^{-r/\alpha_\psi}$ (where $M_0$ is the central mass density and $\alpha_\psi$ is the scale length). General morphological parameters can be obtained by integration of $\psi(t)$. Table \ref{t2} shows some of them, namely the scale length, core radius, and total stellar mass for each galaxy.

In Cetus, most stars are older than $\sim 9$~Gyr regardless of radius. In Tucana, the end of the star formation is also centered around 9 Gyr, but it seems to be shifted toward 7 Gyr for $r\lesssim 200$~pc, pointing to a weak population gradient. These ages are lower limits because of the broadening effect by the observational effects on the SFH \citep{monellietal2010a,hidalgoetal2011}. The BSS populations of Cetus and Tucana mentioned in \S\ref{secsfh} are also observed here. These stars extend beyond two scale lengths with a nearly flat radial distribution, supporting the conclusion by \citet{monellietal2012b} that they are BSSs. In the case of LGS-3, stars older than $\sim 8$~Gyr are present along the full range of covered galactocentric distances. An important population of stars younger than $\sim 8$~Gyr also exist, but they are not present in the outer regions of the galaxy.  

The results shown above point to a relation between the ages of the stellar populations and their radial extension. Strong observational evidence of this relation have been observed in other dwarf galaxies \citep*{battinelli&demers2000,aparicio&tikhonov2000,monellietal2003,battagliaetal2006,hidalgoetal2008,gallartetal2008,kolevaetal2011,zhangetal2012}. In all the cases, the young stellar population is more concentrated toward the center than the older stellar population. The two most likely scenarios proposed to account for this characteristic are: (i) stars were formed in an inner gas envelope with a galactocentric radius significantly different from that seen in the present \citep*[][though these authors refer to higher mass disk galaxies]{roskaretal2008,yoachimetal2012,birdetal2012} and (ii) the stars were formed at a galactocentric radius comparable to that presently observed \citep*{hidalgoetal2003,stinsonetal2009}. In the first case, the stellar population segregation is produced by stellar migration 
or accretion: older stars can reach, on average, larger galactocentric distances than younger stars. In the second one, the stellar population segregation is produced by an outside-in quenching process of the star formation. For simplicity we will call scenario (i) ``migration'' and scenario (ii) 'in situ'.

\begin{deluxetable*}{ccccccc}[h]
\tabletypesize{\scriptsize}
\tablecaption{SFRs, ages, and metallicities as a function of radius.\label{t3}}
\tablewidth{0pt}
\tablehead{
\colhead{Galaxy} &\colhead{r} &\colhead{$<\psi_{e10-e95}>$} &\colhead{$<age_{e10}>$} &\colhead{$\rm <[M/H]_{e10}>$} &\colhead{$<age_{e95}>$} &\colhead{$\rm <[M/H]_{e95}>$} \\
\colhead{} &\colhead{(s.l.u.)\tablenotemark{a}} &\colhead{($10^{-9} M_\sun yr^{-1} pc^{-2}$)} &\colhead{(Gyr)} &\colhead{} &\colhead{(Gyr)} &\colhead{}}
\startdata
CETUS     &       &                &                &                 &              &               \\                                               
\nodata   &$R_1$  &$2.20\pm 0.04$  &$13.2\pm 1.1$   &$-1.74\pm 0.30$  &$9.6\pm 0.2$  &$-1.56\pm 0.20$\\
\nodata   &$R_2$  &$1.47\pm 0.02$  &$13.2\pm 1.1$   &$-1.73\pm 0.29$  &$9.3\pm 0.2$  &$-1.52\pm 0.18$\\
\nodata   &$R_3$  &$0.88\pm 0.01$  &$13.3\pm 1.0$   &$-1.73\pm 0.29$  &$9.3\pm 0.2$  &$-1.45\pm 0.15$\\
\nodata   &$R_4$  &$0.22\pm 0.00$  &$13.3\pm 0.7$   &$-1.67\pm 0.25$  &$9.8\pm 0.2$  &$-1.47\pm 0.16$\\
TUCANA    &       &                &                &                 &              &               \\
\nodata   &$R_1$  &$1.88\pm 0.02$  &$13.2\pm 0.7$   &$-1.66\pm 0.25$  &$8.2\pm 0.1$   &$-1.27\pm 0.10$\\
\nodata   &$R_2$  &$1.56\pm 0.02$  &$13.3\pm 0.6$   &$-1.62\pm 0.23$  &$8.9\pm 0.1$   &$-1.17\pm 0.08$\\
\nodata   &$R_3$  &$1.36\pm 0.01$  &$13.3\pm 0.5$   &$-1.57\pm 0.20$  &$9.9\pm 0.1$   &$-1.51\pm 0.18$\\
\nodata   &$R_4$  &$0.23\pm 0.00$  &$13.3\pm 0.6$   &$-1.62\pm 0.23$  &$10.1\pm 0.1$  &$-1.48\pm 0.16$\\
LGS-3     &       &                &                &                 &              &               \\
\nodata   &$R_1$  &$0.79\pm 0.01$  &$12.8\pm 0.8$   &$-1.68\pm 0.32$  &$1.9\pm 0.0$  &$-1.10\pm 0.09$\\                                               
\nodata   &$R_2$  &$0.67\pm 0.01$  &$12.8\pm 0.8$   &$-1.68\pm 0.32$  &$4.0\pm 0.1$  &$-1.39\pm 0.16$\\
\nodata   &$R_3$  &$0.47\pm 0.01$  &$13.0\pm 0.8$   &$-1.68\pm 0.32$  &$5.6\pm 0.1$  &$-1.49\pm 0.20$\\
\nodata   &$R_4$  &$0.17\pm 0.00$  &$13.2\pm 0.7$   &$-1.65\pm 0.30$  &$7.4\pm 0.1$  &$-1.63\pm 0.28$\\
PHOENIX   &       &                &                &                 &              &               \\
\nodata   &$R_1$  &$1.39\pm 0.01$  &$12.6\pm 0.4$   &$-1.67\pm 0.21$  &$1.4\pm 0.0$  &$-1.08\pm 0.07$\\
\nodata   &$R_2$  &$1.16\pm 0.01$  &$13.0\pm 0.6$   &$-1.70\pm 0.28$  &$4.2\pm 0.1$  &$-1.31\pm 0.15$\\
\nodata   &$R_3$  &$0.77\pm 0.01$  &$13.2\pm 1.2$   &$-1.75\pm 0.30$  &$5.9\pm 0.1$  &$-1.54\pm 0.14$\\
\enddata
\tablenotetext{a}{scale length units.}
\end{deluxetable*}

A detailed analysis of $\psi(t,r)$ may shed light on the different galaxy formation models. The scale length of each stellar population and the age of the stars as a function of radius can be obtained from integration of $\psi(t,r)$. The former will be discussed in \S \ref{secgro}. The latter is obtained by calculating the age corresponding to certain percentiles of $\psi(t)$. The  age of the $p$-th percentile ($e_{p}$) of $\psi(t)$ is defined as the age at which $\Psi(t)=p/100\times\Psi(T)$. We will discuss the ages of the 10th ($e_{10}$) and 95th ($e_{95}$) percentiles, which we consider representative of the age of the first and last star formation events, respectively. The analysis of $e_{10}$ and $e_{95}$ as a function of radius, as described in \citet{hidalgoetal2009}, gives more valuable information on the gradients of stellar populations than the simple mean age of the stars.
Table \ref{t3} shows the mean SFRs, ages, and metallicities for $e_{10}$ and $e_{95}$ for the four galaxies.

\begin{figure}[h]
\centering
\includegraphics[width=9cm,angle=0]{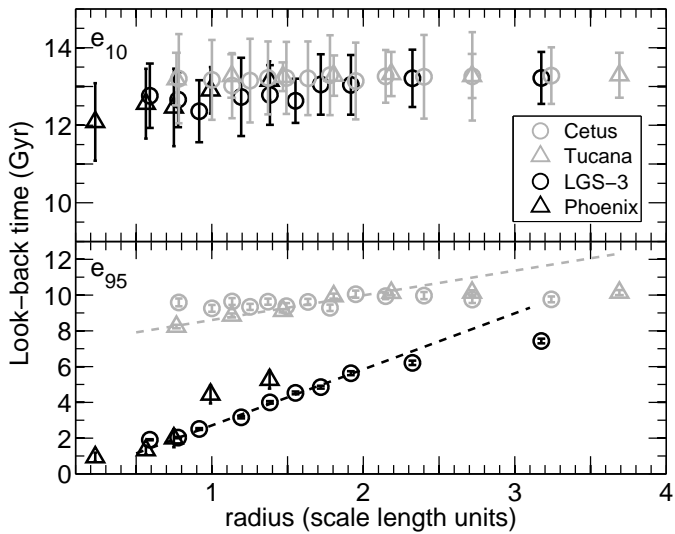}
\protect\caption[ ]{The age of the 10th (top panel) and 95th (bottom panel) percentiles of the cumulative mass function of Cetus, Tucana, LGS-3, and Phoenix as a function of radius.
Two straight lines have been fitted to the inner points of $e_{95}$ ($0.75\alpha_\psi\leq r\leq 2\alpha_\psi$) for Tucana (gray dashed-line) and LGS-3 (black solid-line).}\label{f9}
\end{figure}

Figure \ref{f9} shows $e_{10}$ and $e_{95}$ as a function of galactocentric radius for Cetus, Tucana, LGS-3, and Phoenix. The $e_{10}$ is, within uncertainties, flat for all the galaxies. The weighted mean age of $e_{10}$ is $12.8\pm 0.2$~Gyr for dTrs and $13.1\pm 0.2$~Gyr for dSphs. The dispersion associated to $e_{10}$ ($\sigma=0.2$~Gyr) is indicative of the variation of the age of the first star formation event with radius. In an ``in situ''  scenario this means that within a time period of $\sim 0.4$~Gyr (i.e., $2\sigma$), the star formation was concurrent at all radii for all the galaxies producing a coeval star formation onset. Early works by \citet{mccray&kafatos1987} and \citet{morietal1997}, and more recently by \citet*{elmegreenetal2002} have shown that stellar winds from OB associations can induce new star formation with time scales of the order of 0.1 Gyr for radii up to 300 pc, and could even reach 600 pc for dwarf galaxies. This is compatible with  our Fig.~\ref{f9}, showing that the values of 
$e_{10}$ reach about 3.5 scale lengths which, in the case of Tucana, represents $\sim 700$~pc. On the other hand, the dispersion in $e_{10}$ can not rule out the ``migrating'' scenario; in order for a star formed in the center to reach a distance of 300 -- 600 pc after 0.2 Gyr, only requires a migration velocity of $\rm 2\sim 3~km~s^{-1}$, which is lower than the typical velocity dispersion of a dwarf galaxy \citep{tolstoyetal2009}.  However, \citet{bastian2011} find no evidence for such migration velocities, likely because the conservation of angular momentum prevents stars from migrating far from their original radius.

Focusing now on the last star formation event, Cetus shows $e_{95}$ centered at $9.6\pm 0.3$~Gyr with no dependency with radius. Tucana $e_{95}$ shows a small gradient of $\sim 1.5$~Gyr per scale length unit for stars with $r\lesssim 2\alpha_\psi$. Thus, the star formation in Cetus lasted about 3.5 Gyr ($e_{10}-e_{95}$) and it stopped at the same time (within the $\sim 0.3$~Gyr of dispersion of $e_{95}$) at all radii. A similar pattern is seen in Tucana for radii larger than about two scale lengths. However, in the inner regions of Tucana, $e_{95}$ decreases toward younger ages for $r\lesssim 2\alpha_\psi$, pointing to a star formation which lasted longer ($\sim 1.5$~Gyr). 
This is in fair agreement with the results obtained by \citet{bernardetal2008}, \citet{bernardetal2009}, and \citet{monellietal2012a} from RR-Lyrae star observations which indicate differences in the early evolution of these two galaxies.

For LGS-3 and Phoenix, $e_{95}$ shows a clear stellar population age gradient of about $\sim 4$~Gyr per scale length in the range $r\gtrsim 0.75\alpha_\psi$. It is interesting to note that Tucana and LGS-3 show a change in the profile of $e_{95}$ (Cetus shows no gradient and the radial sampling in the outer regions of Phoenix is insufficient). This change is produced at about two scale lengths for both galaxies and corresponds to ages $\sim10$~Gyr for Tucana and $\sim 6$~Gyr for LGS-3. For Tucana, this is concurrent with the sharp drop in its SFR (See Figs.~\ref{f3} and \ref{f8}).

\begin{figure}[h]
\centering
\includegraphics[width=9cm,angle=0]{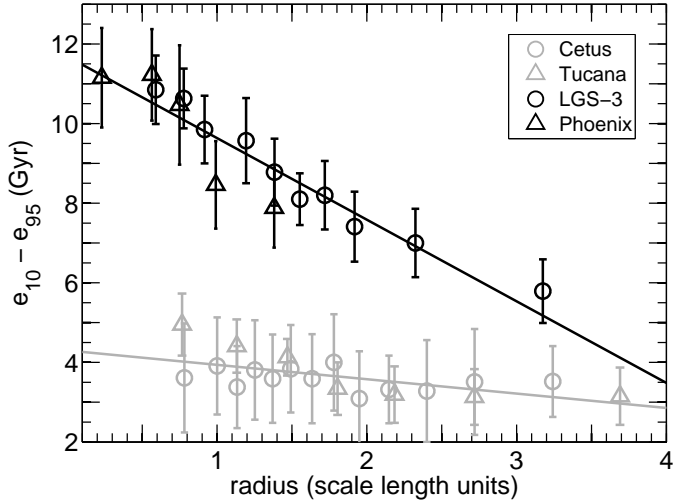}
\protect\caption[ ]{Difference between the ages of the 10th and 95th percentile ($e_{10}-e_{95}$) as a function of radius for the four galaxies.
Two straight lines have been fitted, one to the dSphs Cetus and Tucana (gray line) and another to the dTrs Phoenix and LGS-3 (black line). The intersection of both lines is
at $r\sim 4.3\alpha_\psi$, $e_{10}-e_{95}\sim 2.8$~Gyr}\label{f10}
\end{figure}

The idea that $e_{95}$ of the dTrs and dSphs could converge at large radii suggests that $e_{95}$ could be coeval for both types of galaxies beyond some radius. To find the radius at which all the galaxies of the sample would have the same $e_{95}$, we have plotted in Fig.~\ref{f10} the difference $e_{10}-e_{95}$ as a function of radius for each galaxy. $e_{10}-e_{95}$ embraces 85\% of the total stellar mass formed at each radius, representing the time elapsed to form the main stellar structure of the galaxies. One straight line has been fitted to each morphological type. Both fits have a negative slope, the smaller $e_{10}-e_{95}$, the less the star formation lasts, indicating that the star formation is kept for a longer time toward the center of the galaxy. The two lines intersect at $r\sim 4.3\alpha_\psi$ and $e_{10}-e_{95}\sim 2.8$~Gyr. This result suggests that beyond $r\sim 4.3\alpha_\psi$ the dTrs and dSphs galaxies analyzed here show similar characteristics: (i) their oldest stars are $\sim 
13$~Gyr old ($e_{10}$), (ii) their mean metallicity is below $\rm [M/H]\sim -1.7$ (see Table \ref{t3}), and iii) the spread in the age of the stars is $\sim 2.8$~Gyr, although it could be lower due to the broadening of the main features of the SFH as a result of observational effects \citep{monellietal2010a,monellietal2010b,hidalgoetal2011}.

\section{THE GROWTH OF STRUCTURE OF DWARF GALAXIES}\label{secgro}

In addition to the galaxies presented in this work, clear evidence of stellar population gradients have been observed in dwarf galaxies within a wide range of physical characteristics. Early studies of the resolved stars with HST revealed smooth underlying populations of RGB stars and concentrated populations of young stars in the Local Group and nearby galaxies Sextans~A, Pegasus, Leo~A, and GR~8 \citep{dohm1997, gallagher1998, tolstoy1998, dohm1998}.  Many other studies followed, e.g., Pegasus, DDO210, and Tucana \citep{battinelli&demers2000}, DDO187 \citep*{aparicioetal2000}, DDO190 \citep{aparicio&tikhonov2000,battinelli&demers2006}, DDO165 and DDO181 \citep{hidalgoetal2003} IC10, Leo-A, and Sextans A \citep{magrinietal2003}, NGC6822 \citep*{leisyetal2005, demersetal2006}, NGC3109 \citep{hidalgoetal2008}, the LMC \citep{gallartetal2008} and the SMC \citep{noeletal2009} are some examples. In all cases, the gradients are in the sense that the mean age of the stellar population is younger toward the 
center of the galaxy. This wide range of dwarf galaxy physical characteristics includes differences in total mass, gas, metallicity, velocity dispersion, and environment suggesting that population gradients are intrinsic to dwarf galaxy formation and/or the early evolution process. 
The results presented in the previous section show that the differences in the characteristics of the isolated dwarfs analyzed tend to disappear at large galactocentric distances, suggesting that some physical processes may not affect all galactocentric radii equally. Thus, these processes can play a role in determining the morphological type of a dwarf galaxy and shaping the radial distributions of their stellar populations. 
The UV-background \citep{bullocketal2000,ricotti&gnedin2005,sawalaetal2010} is one physical process that may have affected the inner and outer regions differently. 
In this section we discuss the effect of the UV-background as a function of radius and its relevance to the radial gradients of the stellar populations.

\subsection{The effect of the cosmic UV-background at the outskirts of the galaxies}

The UV-background acts not only by suppressing gas accretion in the lowest mass galaxies but also by heating the remaining gas above the virial temperature of the DM halo and expelling it \citep{bullocketal2000}. \citet{susa&umemura2004}, \citet{ricotti&gnedin2005}, and \citet{sawalaetal2010} have suggested that internal thermal feedback may be necessary prior to the UV-background to shut down star formation. In any case, the final result is that in the lowest mass galaxies, those with total mass $\rm \lesssim 2\times10^8 M_\sun$ at the epoch of reionization (EoR) \citep{barkana&loeb1999}, the star formation should switch off at redshift $z\sim 6$ in a very short time in the whole galaxy. As a result, low mass galaxies would only show star formation during a short time interval. However, in higher mass galaxies which are able to keep dense gas in the center, self-shielding would prevent the UV-background from quenching the star formation \citep{tajiri&umemura1998,sawalaetal2010}, allowing a longer 
formation time interval. 

In low mass galaxies, even if self-shielding is at play, there would be a non self-shielded region where the star formation should have stopped at the EoR. Here, we have performed several tests with mock stellar populations to elucidate whether the SFHs of our galaxies hold some signature of the EoR at some galactocentric radius. We have followed the procedure described in \citet{monellietal2010a,monellietal2010b} and \citet{hidalgoetal2011}, but will apply it to radial samples of the galaxies (as opposed to the entire galaxy). In short, the modeling procedure consists of building a synthetic CMD corresponding to an instantaneous burst of star formation. Observational effects are simulated in the synthetic CMD and the corresponding SFH is recovered in the same way as for the observed CMDs. Assuming a flat Einstein-de Sitter Universe and using the 5-year WMAP data \citep{komatsuetal2009}, the age of the EoR ($z=6$) is 12.7 Gyr \citep{loeb&barkana2001,beckeretal2001}. We will assume this as the age in which 
the Universe is fully reionized.

\begin{figure}[h]
\centering
\includegraphics[width=9cm,angle=0]{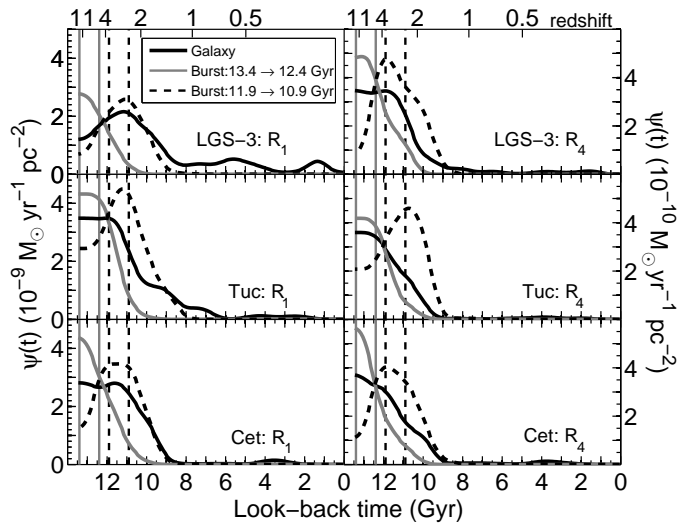}
\protect\caption[ ]{Input (vertical lines) and recovered SFHs for two mock bursts placed before (gray solid-line) and after (black dashed-line) the EoR. Two regions were selected: $R_1$ with stars $r\lesssim\alpha_\psi$ (left panels) and $R_2$ with $r\gtrsim2\alpha_\psi$ (right panels). The SFRs of the galaxies are overplotted (black solid-line).}\label{f11}
\end{figure}

We show the results of two of our tests in Fig. \ref{f11}: one for a starburst starting before the EoR (first burst-model), between 13.4 and 12.4 Gyr ago, and another for a burst produced after the EoR (second burst-model), between 11.9 and 10.9 Gyr ago. We have performed this analysis at regions $R_1$ and $R_4$ defined in \S\ref{secsfh}, which contain stars located at $r\lesssim\alpha_\psi$ and $r\gtrsim2\alpha_\psi$, respectively. The results show that for $R_1$, the first burst-model is inconsistent with the observed SFHs of all of the galaxies. Indeed, the best burst-model for the inner regions of LGS-3 and Cetus is the second burst-model, i.e., one with all stars formed after the EoR. Regarding the $R_4$ region, only for the case of Tucana there is some degree of agreement between the results and the first burst-model, but even in this case, the agreement is not strong. We have to conclude that the UV-background has had little effect, if any, on the SFH of the three galaxies, even at their outermost 
regions. This likely indicates that i) the galaxies in our sample are more massive than the low mass galaxies that are predicted to have been strongly affected by the UV-background or ii) the strong effects of the UV-background were constrained by a self-shielding mechanism that would extend at least up to two scale lengths from the center of the galaxies. In the latter case, imprints of the UV-background might be found beyond the observed regions.

\subsection{The profile of the stellar populations spatial distribution}

In this section we will discuss how the radial distribution of the stellar mass, $\psi(r)$, evolves with time. Temporal slices at any look-back time can be obtained from the general solutions shown in Figure \ref{f8}. Each temporal slice will show $\psi(r)$ for the corresponding look-back time. For Cetus and Tucana, in agreement with the results present in \S\ref{secsfh}, the analysis of the radial distribution of the stellar mass has been limited to stellar populations older than 5 Gyr. For the case of Phoenix, $\psi(r)$ has been obtained at the five elliptical regions shown in Figure \ref{f1}.

\begin{deluxetable*}{cccccc}[h]
\tabletypesize{\scriptsize}
\tablecaption{Scale length as a function of time.\label{t4}}
\tablewidth{0pt}
\tablehead{
\colhead{Galaxy} &\colhead{$t\geq 10$~Gyr}  &\colhead{$5\leq t <10$~Gyr}  &\colhead{$2\leq t< 5$~Gyr} &\colhead{$1\leq t< 2$~Gyr} &\colhead{$t< 1$~Gyr}\\
\multicolumn{6}{c}{$\alpha_\psi$ (pc)}}
\startdata
CETUS      &$226\pm~~\,\,5$         &$169\pm 15$     &\nodata             &\nodata       &\nodata   \\
TUCANA     &$141\pm~~\,\,7$         &$~\,84\pm~\,7$  &\nodata             &\nodata       &\nodata   \\
LGS-3      &$165\pm~\,  16$         &$~\,81\pm~\,6$  &$~\,83\pm~\,7$      &$83\pm  3$    &$55\pm  4$\\
PHOENIX    &$526\pm    293$         &$205\pm 31$     &$110\pm~\,  5$      &$68\pm  8$    &$70\pm  8$\\
\enddata
\end{deluxetable*}

\begin{figure}[h]
\centering
\includegraphics[width=9cm,angle=0]{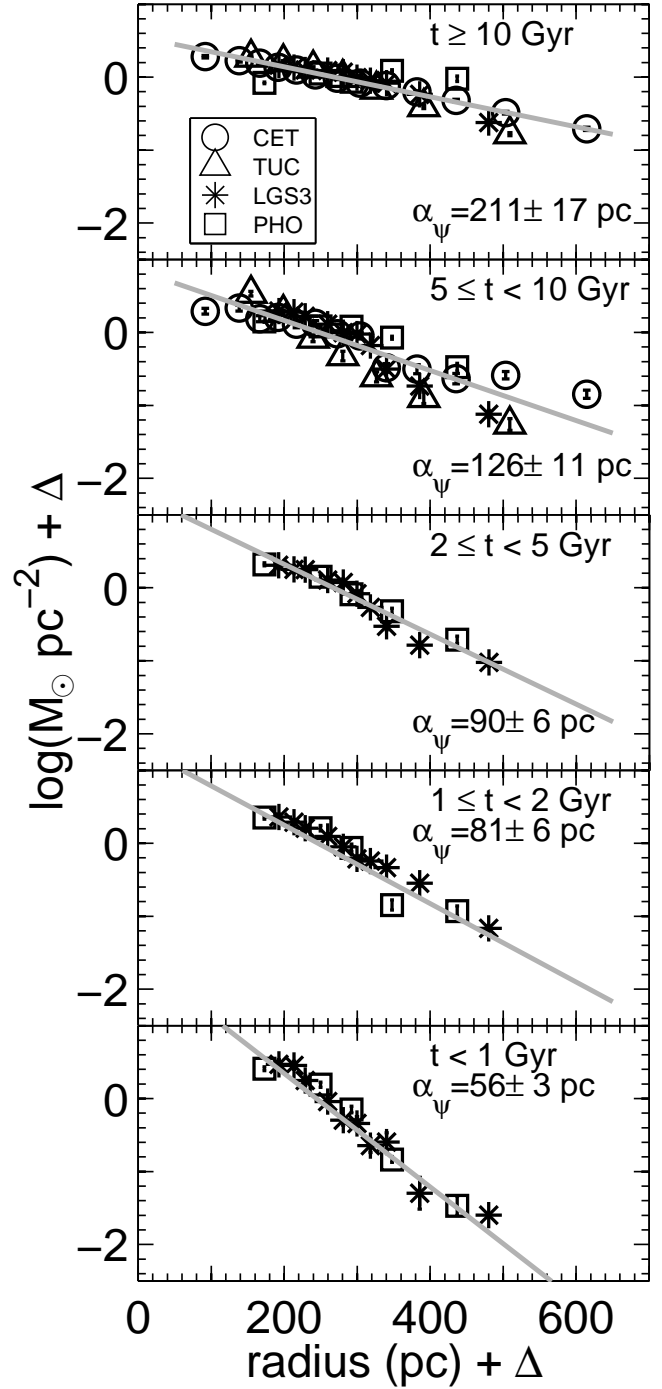}
\protect\caption[ ]{Radial mass distribution $\psi(r)$ of Cetus, Tucana, LGS-3, and Phoenix for five time periods (older stars in top panel). For each galaxy and time period, $\psi(r)$ has been recentered using the barycenter of the distribution (which has been denoted with $\Delta$ in the axes). An exponential fit (gray line) to all the galaxies has been used to obtain the scale length $\alpha_\psi$ of the radial distribution, which is shown in each panel. An overall decrease of  $\alpha_\psi$ with time can be observed.}\label{f12}
\end{figure}

Figure \ref{f12} shows $\psi(r)$ for each galaxy and five specific periods of time. For a direct comparison between all of the galaxies, at each time period, $\psi(r)$ has been re-centered by using the barycenter of the distribution of each galaxy. For each time period, an exponential fit to all $\psi(r)$ provides the mean scale length $\alpha_\psi$. The overall behavior of $\alpha_\psi$ is to decrease, with time, from the epoch of the formation of the galaxy to the present. However, three epochs can be distinguished: i) old ($t\geq 10$~Gyr) where $\alpha_\psi$ attains its largest value, ii) intermediate ($1\leq t<10$~Gyr) where $\alpha_\psi$ is $\sim 50\%$ of its largest value but it remains nearly constant for a long time, and iii) young ($t<1$~Gyr) where $\alpha_\psi$ drops to its lowest value which is a $\sim 30\%$ of the old one. In other words, in all four galaxies, the oldest stellar populations are more spatially extended than the younger ones.

The change in $\alpha_\psi$ with time is also observed in each single galaxy, as can be seen in Table \ref{t4}. It shows $\alpha_\psi$ for each time period and galaxy. For all the galaxies, $\alpha_\psi$ decreases from old to young ages. The overall decrease of $\alpha_\psi$ with time could be the result of either or both the 'in situ' or 'migration' scenarios introduced in \S\ref{secrad}. However, \citet{stinsonetal2009} and \citet{schronyenetal2013} have studied models of evolution of low mass galaxies in which stellar radial migration shows no large impact on the final radial distribution of stars. If this were the case, the 'in situ' scenario would be predominant. The decrease of $\alpha_\psi$ with time would be the result of the quenching of the star formation towards the center when the galaxy is running out of gas in the outskirts, i.e., an outside-in scenario for star formation. Kinematics and chemical abundances of individual stars as a function of radius could shed some light on the true 
origin of the correlation of $\alpha_\psi$ with time.

\section{SUMMARY AND CONCLUSIONS}\label{secsum}
                                                                                                                                                      
We have analyzed the SFHs as a function of the galactocentric radius of four isolated dwarf galaxies of the Local Group. The results presented in this paper will be summarized in the following.

The oldest stellar populations of the dSphs and dTrs in our sample are, within the errors, coeval ($\sim 13$~Gyr) at all galactocentric radii. The dTrs show a prominent radial gradient in their measured SFHs in the sense that the younger populations are more centrally concentrated; the centers of these galaxies host young ($\lesssim 1$~Gyr) populations while the age of the last formation event increases smoothly with radius at a rate of about $\sim 4$~Gyr per scale length unit, beyond 0.75 scale lengths. This contrasts with the dSph galaxies, where Tucana shows a small gradient of $\sim1.5$~Gyr per scale length unit, within the two central scale lengths, while no measurable gradient is found in Cetus.

The differences in the CMDs and in fundamental properties in our sample of dwarf galaxies, such as the SFHs and the AMRs, tend to disappear as the galactocentric radius increases. These results suggest that, regardless of the morphological type, these dwarf galaxies show common characteristics for galactocentric radii $r\gtrsim 4\alpha_\psi$. The stellar content of this region would be composed of a very old, metal-poor stellar population. Our results are compatible with a scenario in which the UV-background at the EoR did not stop the star formation in the inner regions of the galaxies ($r\lesssim 2\alpha_\psi$) and had little effect, if any, in the external ones. Within the innermost region, self-shielding may have played a role in preserving the gas and star formation activity.

Different behaviors have been found in the metal enrichment processes of the galaxies. The dTrs show two phases of metallicity enrichment. A first phase, accounting for $\sim 85\%$ of the total mass of stars formed, shows virtually no metallicity enrichment. A second phase, producing the remaining $\sim 15\%$ of the total mass, shows a relatively sharp metallicity enrichment and at lower SFRs than in the first phase. Comparison with simple chemical evolution models suggest that a simple explanation can be achieved by assuming that in the first phase, the stars are formed mostly from fresh gas with enriched material ejected from the galaxy. In the second phase the stars would be formed mostly from gas enriched from stellar ejecta.

In the three galaxies in which we have measured significant stellar population gradients, we have determined that the exponential scale lengths of the radial distributions of the stellar populations decrease with time. These results are in agreement with outside-in scenarios of dwarf galaxy evolution, in which a quenching of the star formation toward the center is produced as the galaxy runs out of gas in the outskirts.

We thank the anonymous referee for the comments and suggestions that helped us to improve this paper. Support for this work was provided by NASA through grant GO-10515 from the Space Telescope Science Institute, which is operated by AURA, Inc., under NASA contract NAS5-26555, the IAC (grant 310394), and the Education and Science Ministry of Spain (grants AYA2004-06343 and AYA2007-3E3507). This research has made use of NASA's Astrophysics Data System Bibliographic Services and the NASA/IPAC Extragalactic Database (NED), which is operated by the Jet Propulsion Laboratory, California Institute of Technology, under contract with the National Aeronautics and Space Administration.

\end{document}